\documentclass[aip,jcp,reprint]{revtex4-1}
\pdfoutput=1
\usepackage{amsmath}%
\usepackage{amsfonts}%
\usepackage{amssymb}%
\usepackage{graphicx}

\usepackage[T1]{fontenc} 
\usepackage[cp1250]{inputenc}

\usepackage{bbm}

\providecommand{\norm}[1]{\lVert#1\rVert}

\newcommand{\adnote}[3]{\makebox[0pt][r]{\makebox[#1][l]{\raisebox{#2}{#3}}}}

\newcommand{\eg}{\textit{e.g.\ }}
\newcommand{\ie}{\textit{i.e.\ }}

\newcommand{\kB}{k_{\rm B}}
\newcommand{\DE}{\Delta \! E}
\newcommand{\eAt}{e\text{\AA}\!{}^3}

\newdimen\mycolumnwidth
\begin{document}
	
\title{Nanoscale ordering of planar octupolar molecules for nonlinear optics at higher temperatures}

\author{Micha{\l} Jarema}
\email{michal.jarema@pwr.edu.pl}
\affiliation{Department of Semiconductor Materials Engineering, Wroc{\l}aw University of Science and Technology, Wybrze\.{z}e Wyspia\'{n}skiego 27, 50--370 Wrocław, Poland}
\author{Antoni C. Mitu\'s}
\affiliation{Department of Theoretical Physics, Wroc{\l}aw University of Science and Technology, Wybrze\.{z}e Wyspia\'{n}skiego 27, 50--370 Wrocław, Poland}
\author{Joseph Zyss}
\affiliation{LUMIN Laboratory and Institut d'Alembert, Ecole Normale Sup\'{e}rieure Paris-Saclay, CNRS, Universit\'{e} Paris-Saclay, 4, avenue des Sciences, Gif-sur-Yvette, France}

\date{\today}
	
\begin{abstract}
We develop scenarios for orientational ordering of an  in-plane system of small flat  octupolar molecules at the low-concentration limit, aiming towards  nonlinear-optical (NLO) applications  at room temperatures. The octupoles interact with  external electric poling fields and intermolecular interactions are neglected. Simple statistical-mechanics models are used to analyze the orientational order in the  \textit{very weak poling} limit, sufficient for retrieving the NLO signals owing to the high sensitivity of NLO detectors and measurement chains. Two scenarios are discussed. Firstly, the octupolar poling field is imparted by a system of point charges; the setup is subject to  cell-related constraints imposed by mechanical strength and dielectric breakdown limit. The very weak octupolar order of benchmarking TATB molecules is shown to emerge at Helium temperatures. The second scenario addresses the \textit{dipoling} of  octupolar molecules with a small admixture of electric dipolar component. It requires a strong field regime to become effective at Nitrogen temperature range. An estimation of the nonlinear susceptibility coefficient matrix for both scenarios is done in the high-temperature (weak interaction) limit formalism. We argue that  moderate modifications of the system like, \emph{e.g.}, an increase of the size of the octupole, accompanied by dipole-assisted octupoling, can increase the poling temperature above Nitrogen temperatures.
\end{abstract}

	
\maketitle 
	

\section{Introduction}
	
Organic molecules and materials have been of persistent interest throughout decades towards the exploration of nonlinear optical (NLO) phenomena and their progress towards  applications.\cite{BoydNLO,KielichNLO,chemla1987nonlinear,MaroulisAtomicmolecularnonlinear2011,messier2012organic} The inherent tensorial properties at all scales promote symmetry considerations at the core of
molecular nonlinear optics alongside propagative and quantum issues. Advances in nanoscale science and technologies have come to enable nonlinear optical configurations all the way from the wavelength scale of waveguided optics and microresonators,\cite{LafargueLocalizedlasingmodes2014} down to the nanoscale \cite{HajjElectroopticalPockelsscattering2011,CastagnaNanoscalePolingPolymer2013,BrasseletNanoCrystalsQuadraticNonlinear2010} and single molecule experiments.\cite{PeyronelQuantumnonlinearoptics2012a} Quadratic NLO processes require centrosymmetry breaking at the scales from individual molecules to bulk interactions in molecular crystals.\cite{ZyssRelationsmicroscopicmacroscopic1982,ZyssChiralityhydrogenbonding1984} Polar conjugated molecules provide a versatile template, moreover embedded in the broader pool of multipolar non-centrosymmetric systems whereby octupolar molecules \cite{ZyssFirst} are a special case of major interest. Multipolar molecules and materials feature richer tensor potential towards more advantageous nonlinear polarization schemes, such as octupolar light--matter configurations abiding to polarization independence conditions.\cite{ZyssJCP93engineeringImplications,BZ98} Due to the symmetry induced net cancellation of their dipole moment that forbids classical dipolar coupling schemes, octupolar molecules have set a challenge since the early stage of their development. Therefore, the search and demonstration of efficient acentric orienting schemes for octupolar molecules remains an active domain of research to this day where theoretical modelling are spurring experiments and vice versa in a currently widely open context.

The first steps towards the evaluation of the required conditions for ordering of low-concentration octupolar molecules by electric field  at nano-scale (nano-octupoling) were reported in Ref.\ \onlinecite{MitusPZ1}. A lattice system of planar two dimensional (2D) molecules with a single in plane rotational degree of freedom was investigated under the assumption of negligible molecular drift motion (\ie fixed molecules restrained to rotate within their plane around their center of mass). In this study, the electric poling field was imparted by a system of electrodes. The molecular interactions as well as the influence of the polymer matrix  on the ordering dynamics were neglected. It was found that effective poling of small octupolar molecules (octupoles)  demanded irrealistic conditions, in the sub-Helium milikelvin temperature range. Such limitation was shown to result from two effects, namely the too small value of a geometric parameter (typical ratio of the sizes of the molecule and poling cell) as well as the spatially inhomogeneous orientation in the lowest energy state of the system. The latter issue was addressed
\cite{ACMNloQo,ActaPhysPolB} by proposing an optimal symmetry adapted poling potential configuration of pure octupolar symmetry that matches the symmetry of the molecular species to be poled. However, this approach failed to lead to a significant increase of the poling temperature. Nevertheless, the aforementioned  studies were just a starting point that relied on a heuristic model with \emph{a priori} estimates of the relevant geometric and physical poling parameters.
	
\begin{figure}
\raisebox{1cm}{\makebox[0pt][l]{a)}}%
\raisebox{-0.58\height}{\includegraphics[height=2.6cm]{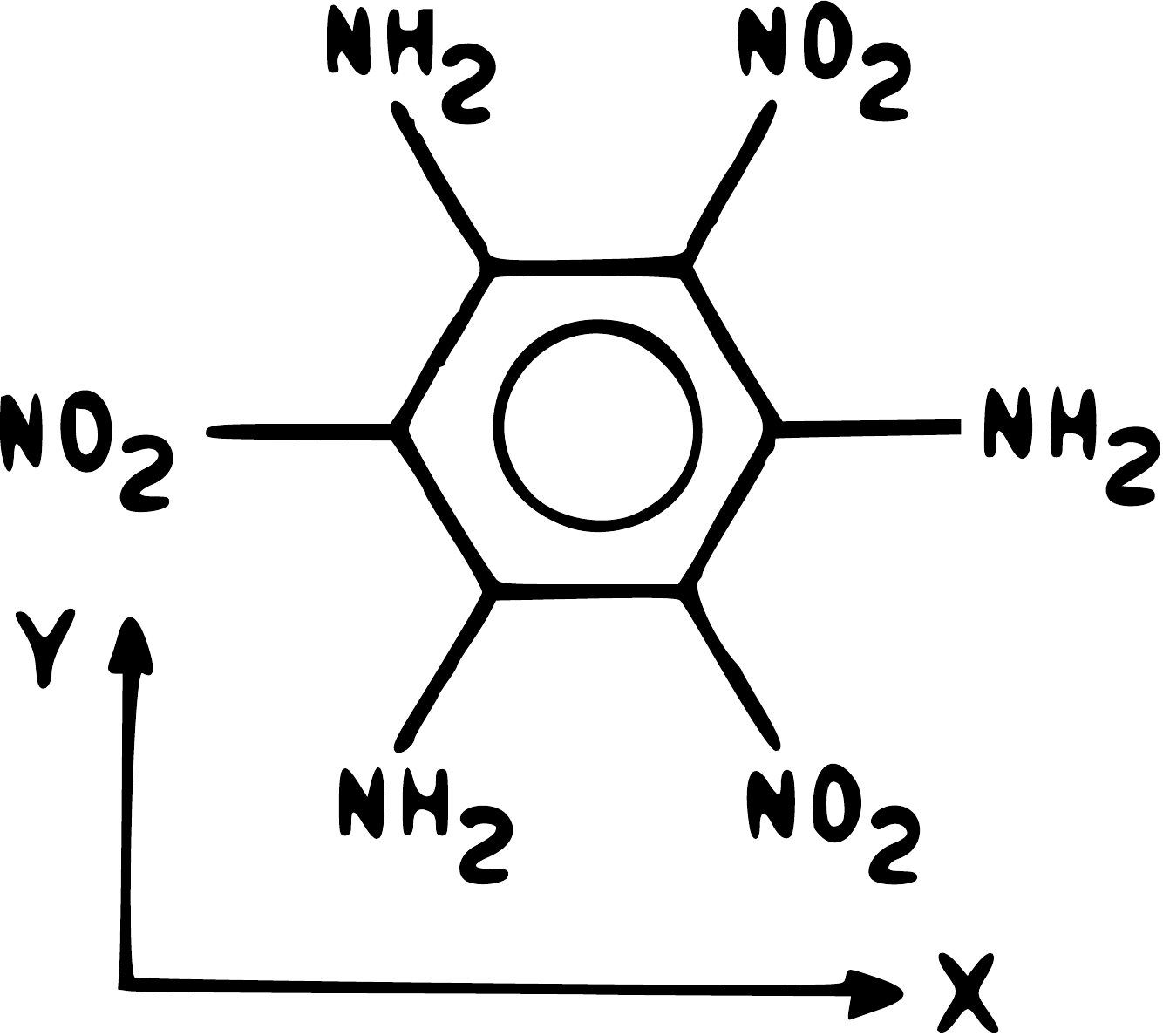}}%
\hspace*{1cm}%
\raisebox{1cm}{\makebox[0pt][l]{b)}}%
\raisebox{-0.5\height}{\includegraphics[height=2.0cm]{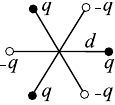}}%
\caption{(a) An exemplary small octupolar molecule: TATB \cite{TATB-SHG-1990} and (b) the model six-arm octupole.
}
\label{fig:TATB-xy}
\end{figure}

\begin{figure}
\centering%
\includegraphics[width=4cm]{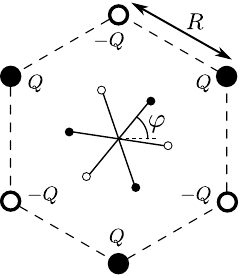}%
\caption{Octupoling setup: the model octupole with orientation angle $\varphi$ at the center of poling cell.
}
\label{fig:poling-cell-charges}
\end{figure}	
	
The objective of this paper is to critically revise the requirements for effective ordering of octupolar molecules in the context of their NLO applications at higher temperatures. This includes (i) statistical-mechanics modeling of poling setups, (ii)  estimation of available values of various physical parameters  and (iii), evaluation of the amplitudes of NLO signals.
	
The paper is organized  as follows. A classical model for an octupole is proposed and discussed in Sect.~\ref{sect:molecule}. The poling setup and the corresponding electrostatic coupling energy are introduced in Sects.~\ref{s:poling-setup-limitations} and \ref{s:Energy}. Statistical-physics aspects of octupoling (order parameters and the concept of very weak octupoling) are introduced and discussed in Sects. \ref{sect:order_par} and \ref{sect:weak_poling}, respectively;  octupoling conditions are extensively analyzed in Sect.\ \ref{s:polingconditions}. Another scenario -- dipoling of  octupolar molecules --  is discussed in Sect.\ \ref{sect:dipoling}. Finally, the resulting magnitude of NLO signals is studied in Sect.~\ref{s:NLO-properties}. A Quantum Chemistry (QChem) based validation of the classical model of octupoling proposed in this study is presented in the Appendix.
	
\section{Octupolar molecule template}
\label{sect:molecule}
	
We model the molecular charge density distribution by point-charge extended multipoles of finite size constructed from the minimal number of point charges that is required to account for the given multipolar symmetry. The use of idealized point multipoles of infinitesimal size might be an over-simplification \cite{ExtededDipolesACMGPIRFK08} for separation distances comparable to the size of a molecule.	There are two main model 2D octupoles\cite{ZyssJCP93engineeringImplications} with three-fold axial symmetry: a three-arm octupole considered previously \cite{MitusPZ1} and a six-arm octupole (6AO) shown in Fig.~\ref{fig:TATB-xy}. The latter consists of six alternating charges $\pm q$ at ends of six arms of length $d$. The angle $\varphi$ between the $x$ axis and a positively charged arm specifies the orientation of octupoles in 2D space. We focus on 6AO molecular template in view of its suitability to accommodate intermolecular interactions.\cite{MJPhD} We point out that the model molecules are treated as rigid and non-polarizable (see  Section \ref{s:Energy} and  Appendix).

The dipole and quadrupole moments of 6AO vanish, while the Cartesian components of its octupole moment are defined by
	\begin{align}
	\mathcal{O}_{ijk}=
	\sum_{n=1}^6 q_n \, (\vec{r}_n)_i \, (\vec{r}_n)_j \, (\vec{r}_n)_k,
	\label{eq:octupoleCartDef}
	\end{align}
where  $q_n$ and $\vec{r}_n$ denote the charge and position of the $n$-th point charge in 6AO molecule. 	The octupole moment is a symmetric Cartesian tensor and is irreducible.\cite{Jerphagnon} At the $\varphi=0$ orientation  most of Cartesian components vanish, except for  $\mathcal{O}_{xxx}=-\mathcal{O}_{xyy}=-\mathcal{O}_{yxy}=-\mathcal{O}_{yyx}=\tfrac{3}{2}qd^3$. Unlike the point octupole, the extended octupole template sustains also multipole moments of order higher than octupolar.

The norm of the octupole moment tensor expressed in Cartesian basis takes the following expression
	\begin{equation}
	\|\mathcal{O}\|=\sqrt{ \sum_{i,j,k} \left(\mathcal{O}_{ijk}\right)^2 }
	=3qd^3.	\label{eq:octupoleNorm}
	\end{equation}
Based on the calculation of the octupole moment for a representative octupolar molecule TATB  (see Appendix) we use the following values of molecule's parameters: size $d=2.44$~\AA\ and molecular partial charge $q=0.66\,e$, where $e$ is the elementary charge.
	
\section{Electric field induced nano-octupolar order}\label{s:Noninteracting}
	
\subsection{Model of point charges octupoling}
	
\subsubsection{Poling setup} \label{s:poling-setup-limitations}
	
The octupolar field distribution is generated by a surrounding set of point  charges. Figure~\ref{fig:poling-cell-charges} shows the octupoling cell  consisting of six point charges of alternating signs $\pm Q$, which are located at the vertices of a regular hexagon of side $R$. The values of $R, Q$, limited by nano-scale fabrication technology and breakdown mechanisms, will be discussed in Section \ref{s:polingconditions}. This poling scheme  differs  from the classical electrode poling cell.\cite{MitusPZ1}
	
\subsubsection{Energy} \label{s:Energy}
	
The potential energy $E(\vec{r},\varphi)$ of the 6AO model molecule  rotated by the angle $\varphi$  and subsequently translated from the cell center by a $\vec{r}$ vector is a sum of
$6\times6$ Coulomb interactions between  molecular and  poling charges. In particular, the energy \textit{$E(\vec{r}=0,\varphi)$} of a molecule located
at the cell's center is given by
	\begin{align}
	\label{eq:pot-ene-charges}
	E(\varphi)&= 6\sum_{m=0}^{5}
	\dfrac{(-1)^{m}\, k_0 \, q \, Q }{\sqrt{d^2+R^2-2dR\cos\left(\varphi - (\frac{1}{2}+m)\frac{\pi}{3}  \right)}}, 
	\end{align}
where the denominator is the distance between a selected poling charge and $m$-th molecular charge, and $k_0$ is the Coulomb constant.
Taylor expansion with respect to the small parameter $d/R \approx 10^{-2}$ (Sect.~\ref{s:polingconditions}) reads:
	\begin{align}
	\label{eq:pot-ene-charges-series}
	E(\varphi)&=\frac{\DE}{2}
	\sin3\varphi + O\big((d/R)^{9}\big),
	\end{align}
where
	\begin{align} 		\label{eq:barrier6QOct}
	\DE  \simeq  45\,\frac{ k_0\,q\, Q }{R}\left(\frac{d}{R}\right)^3
	=
	15  \,\frac{ k_0\, Q }{R^4} \,\norm{\mathcal{O}}
	\end{align}
denotes the maximum--minimum energy difference and plays the role of energy barrier for a molecule at the cell center, oscillating around its ground state orientation $\varphi_0=\pi/2$. Thus, in the first approximation  $E(\varphi) = \sin3\varphi$, as for the case of purely octupolar potential.\cite{ACMNloQo} We conclude that the energy barrier
$\DE$ at the center of the poling cell is proportional to the octupole moment of 6AO, and to the cell-related factor ${Q}/{R^4}$ that characterizes the amplitude of the octupoling field.

To test the reliability of a purely classical description of the molecule and its interaction with the poling electric field, we have calculated the energy of TATB molecule at the center of the  poling cell using quantum chemistry simulations (Gaussian98, B3LYP/cc--pVDZ). We have found that the energy barrier agrees well (within 2\%) with our classical model (see Appendix for more details).
	
\subsubsection{Ground state}

\begin{figure}
	\centering
	\newdimen\w
	\w=\mycolumnwidth
	\includegraphics[width=0.49\w]{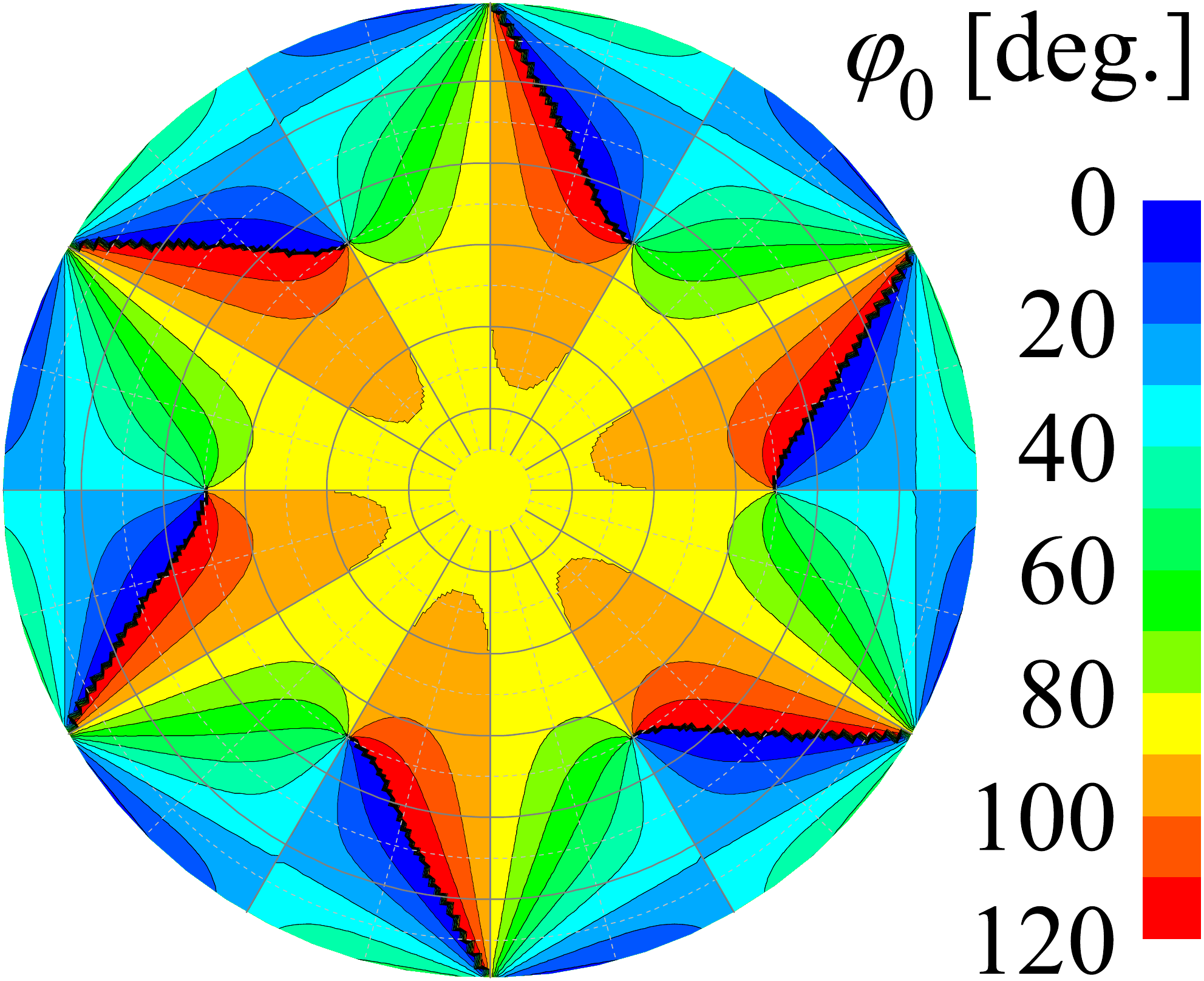}\hspace{0.5cm} 
	\includegraphics[width=0.49\w]{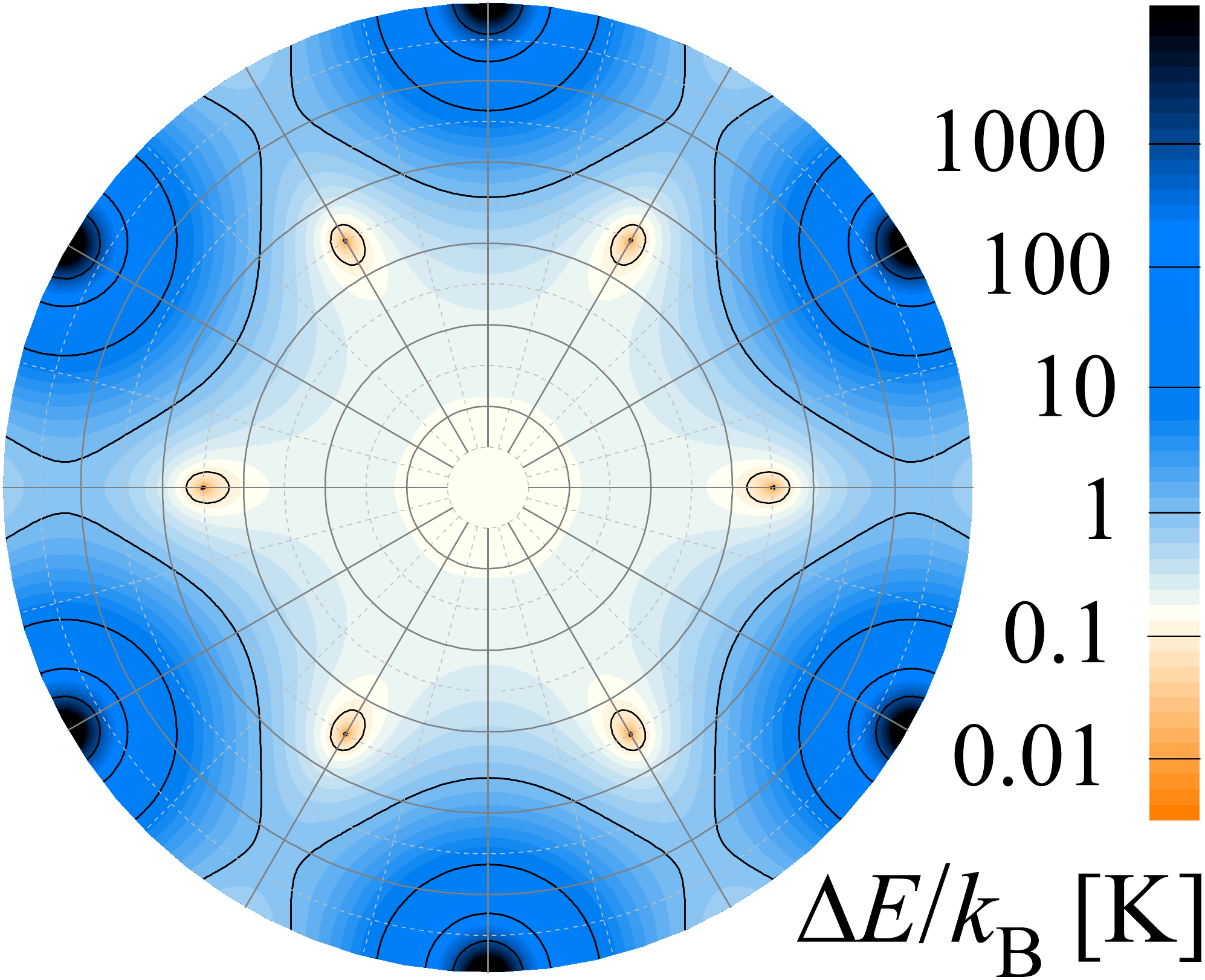}%
	\caption{Ground state orientation $\varphi_0(\vec r)$ (left) and the energy barrier $\Delta E(\vec r)/k_B$ (right) in the six point charges poling cell, \textit{cf.\ } Fig.~\ref{fig:poling-cell-charges}. The  scale of the energy barrier was calculated for $Q=300$~e and $R=60$~nm. }
	\label{fig:6QGS-orientation}
\end{figure}

Figure~\ref{fig:6QGS-orientation} shows the ground state orientation $\varphi_0(\vec r)$ and the energy barrier
$\Delta E(\vec r)/k_B$ in the poling cell ($k_B$ denotes Boltzmann constant).
Those functions have been calculated analytically by direct evaluation of $E(\vec{r},\varphi)$, in contrast to Ref.~\onlinecite{MitusPZ1}, where Monte Carlo simulations were used. 	The ground state orientation is inhomogeneous since the poling potential differs from the pure octupolar potential.\cite{ActaPhysPolB} Nevertheless, in a wide central region, say $r<R/2$, the poling conditions are approximately homogeneous: $\varphi_0(\vec{r})=90^{\circ}\pm10^{\circ}$. On leaving the center and approaching the cell boundary deviations from an homogeneous order become significant. The vortexes (\ie points where the orientation is indefinite due to vanishing energy barrier) are located at distance $r\approx 0.58R$, similar to that in the electrode poling cell.\cite{MitusPZ1} Detailed investigation of this interesting feature falls beyond the scope of this paper and will be undertaken later.

\subsection{Octupoling conditions: statistical mechanics analysis}
	
\subsubsection{Order parameter and its temperature dependence}
\label{sect:order_par}

The local orientational order parameter for a two-dimensional molecule with $n$-fold symmetry axis perpendicular to the plane where molecules are constrained to rotate can be conveniently defined as
\begin{equation}
    \label{eq:pJ}
	p_n(\vec{r_k})= e^{i n \varphi_k},
	\end{equation}		
where $\vec{r_k}$, $\varphi_k$ denote respectively the center of a $k$-labeled molecule and its orientation.  For example, $n=1$ and $n=6$ correspond to a dipole and 2D hexagon,\cite{PatashinskiNanofluidicManifestationsStructure2019} respectively. A flat 6AO octupole has  the three-fold symmetry ($n=3$); thus the corresponding order parameter reads \cite{MitusPZ1}:
\begin{equation}
	p_3(\vec{r_k}) = e^{3i\varphi_k}.
\end{equation}		

The canonical average ${P_n}(\vec{r})$ of $p_n(\vec{r})$ is referred to as the average local order parameter:
	\begin{equation}
	{P_n}(\vec{r}) = \left\langle p_n(\vec{r}) \right\rangle = \frac{1}{Z}\,\int_0^{2 \pi/n} p_n(\vec r) e^{-\beta E(\vec r, \varphi)} {\,\rm d} \varphi,
	\label{eq:Pav}
	\end{equation}
where $\beta=1/\kB T$ ($T$ is an absolute temperature) and $Z$ denotes the partition function:
\begin{equation}
Z = \int_0^{2 \pi/n} e^{-\beta E(\vec r, \varphi)} {\,\rm d} \varphi.
\end{equation}
We will skip thereafter (unless otherwise stated) the index 3 for octupolar order parameters, thus $p = p_3$ and $P = P_3$. The overall order for $N$ octupoles in the system is described by the average global order parameter
	\begin{equation}
	{\mathbb P} = \frac{1}{N}\sum_{i=1}^{N} P(\vec{r}_i).
	\label{eq:Pavglobal}
	\end{equation}
	
Let us estimate the degree of average local octupolar order at the center of the cell, see Fig. \ref{fig:6QGS-orientation}. We define the dimensionless inverse temperature
	\begin{equation}\label{eq:beta*}
	\beta^\star = \frac{\DE}{2}\frac{1}{ \kB\, T} =
	\frac{15\, k_0\, Q \, \norm{\mathcal{O}}}{2\, R^4\,k_B\,T},
	\end{equation}
and use the first term in the expansion of the energy, Eq.\ \eqref{eq:pot-ene-charges-series}. Then,  the partition function is
	\begin{equation}
	Z = \int_{0}^{2\pi/3}\!  e^{- \beta^\star\sin3\varphi} {\,\rm d}\varphi = \frac{2\pi}{3} I_0(\beta^\star),
	\end{equation}		
where $I_n$ stands for the modified Bessel function of the first kind \cite{BesselWatson} for $n=0,1,2,...$.
The average local order parameter reads
	\begin{align}
	\begin{split}
	 P &= \langle p \rangle = \langle e^{3 i \varphi} \rangle = \langle \cos3\varphi\rangle + i\langle \sin3\varphi\rangle
	\\
	 = 0 &+ i \frac{1}{Z} \int_{0}^{2\pi/3} \!\!\!\!\!\!\! \sin3\varphi \, e^{- \beta^\star\sin3\varphi} \, {\,\rm d}\varphi
	= I_{1,0}(\beta^\star)\,e^{3i\pi / 2},
	\end{split}\label{eq:avg-ord-par}
	\end{align}
where $I_{1,0}=I_1/I_0$. The fluctuations of the complex local order parameter $p(\vec r)$, represented by variances of its real ($\Re\,p$) and imaginary ($\Im\, p$) parts, read
	\begin{align}
	\begin{split}
	 {\rm Var}\, \{\Re\,p\} &= \langle(\Re\,p)^2 \rangle - \langle\Re\,p\rangle^2 = I_{1,0}(\beta^\star) / \beta^\star, \\
	 {\rm Var}\, \{\Im\, p\} &=\langle(\Im\,p)^2 \rangle - \langle\Im\,p\rangle^2 = 1 - I_{1,0}(\beta^\star)^2  -  I_{1,0}(\beta^\star) / \beta^\star.
	\end{split}\label{eq:avg-ord-par-variance}
	\end{align}
	
\begin{figure}
		\centering
		\includegraphics[width=\mycolumnwidth]{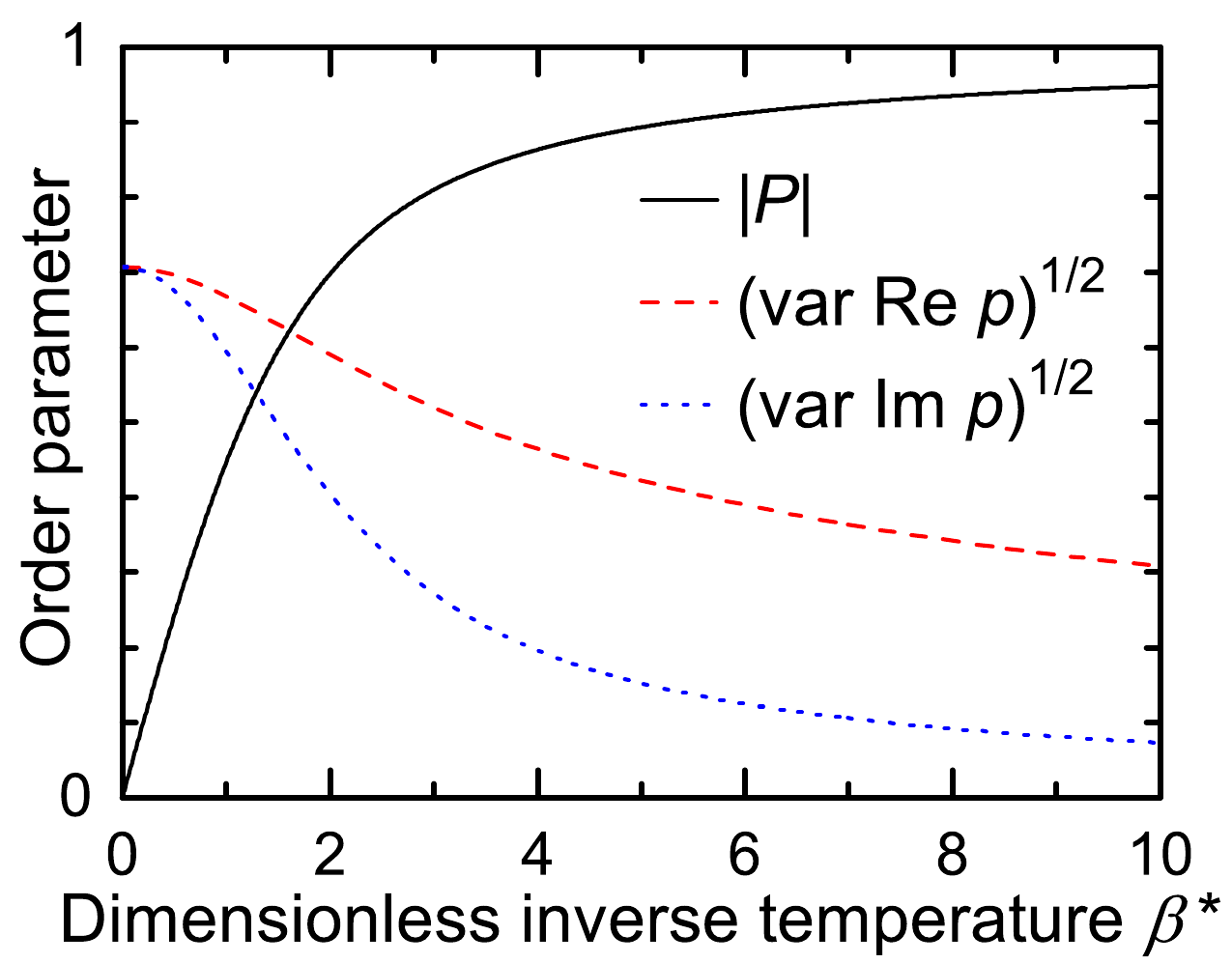}%
		\caption{Characterization of local order $p$ as function of dimensionless inverse temperature $\beta^\star$:  $|P|$ (black line),  $\sqrt{{\rm Var}\, \{\Re\,p\}}$ (red dashed line) and $\sqrt{{\rm Var}\, \{\Im\,p\}}$ (blue dotted line).}
		\label{fig:P-and-VarP-vs-beta}
\end{figure}

Figure \ref{fig:P-and-VarP-vs-beta} shows the plots of $|P|, \sqrt{{\rm Var}\, \{\Re\,p\}}, \sqrt{{\rm Var}\, \{\Im\,p\}}$ as a function of $\beta^\star$. For $\beta^\star \to \infty$ (low temperature)  $P \to -i$, \ie $|P| \to 1$.
At low temperatures the amplitude $|P|$ is close to 1, but the phase still fluctuates.

\subsubsection{Very weak poling regime: energy barriers and experimental conditions}
\label{sect:weak_poling}
	
Consider a system with a rotational degree of freedom in an external poling field  characterized by a potential well with energy barrier $\Delta E$. As long as  thermal fluctuations are much lower than the energy barrier, \ie $\kB T  \ll \DE$, the system is orientationally ordered.  For $\kB T  \approx \DE$ the system is still moderately ordered but strong fluctuations come in. Finally, in the case when  $\kB T  \gg \DE$ the fluctuations become completely disordering. Quite surprisingly, some NLO experimental effects, that directly relate to an average orientational order,  can still be observed when the non-centrosymmetric order is negligible because of large fluctuations, thanks to the sensitivity and high signal-to-noise ratio of current detectors such as photomultipliers or photodiodes. Therefore, in parallel with the strong poling regime ($\kB T  \ll \DE$) and weak poling regime ($\kB T  \approx \DE$) we introduce the \emph{very weak poling regime}  ($\kB T  \gg \DE$), in which residual ordering can still be detected in current NLO experiments. In what follows we estimate the order of magnitude of corresponding (minimal) acentric order in the case of  one of the most important NLO poling experiments that is the electric field induced second harmonic generation (EFISH).\cite{LevineSecondthirdorder1975,OudarOpticalnonlinearitiesconjugated1977}

The typical experimental conditions of EFISH in the case of simple dipolar push--pull molecules are as follows. 	The homogeneous poling electric field strength is of the order of dielectric strength of air $\mathcal{E}_\text{air}\approx3\cdot10^6\;\frac{\rm V}{\rm m}$.	The ground-state dipole moment of dipolar push--pull molecules used in NLO is of the order of 10~D. For example, for pNA molecules with a dipole moment $\mu=6$~D  \cite{DipoleMoments,Sinhagroundexcitedstate1991} the energy barrier reads
	\begin{align}
	\DE_\text{EFISH} &= 2\, \mu \,\mathcal{E}_\text{air} \approx 0.75 \text{ meV},
	\label{eq:EFISHBarrier}
	\end{align}
which compared to the the amplitude of thermal fluctuations $k_B T$ at room temperature ($T=300$ K) yields a factor $\alpha$
	\begin{align}
	\alpha = \frac{\kB\,300\,{\rm K}}{\DE_\text{EFISH}} &\approx {30}.
	\end{align}
The barrier is much lower than the amplitude of thermal fluctuations, therefore the acentric order in EFISH is very low from a statistical-mechanics point of view. Namely, in the two-state model the degree $|m|$ of the orientational  order is
\begin{equation}
\label{eq:two-state}
|m| = \tanh \frac{\mu \mathcal{E}_\text{air}}{k_B T} = \tanh \frac{1}{2 \alpha} \approx 1.6\cdot 10^{-2},
\end{equation}
which appears to be is sufficient for SHG detection.

Since the fundamental requirement of centro-symmetry breaking applies to both EFISH and octupoling, we expect that SHG due to the acentric order in a system of octupoles can occur at a similarly low magnitude of the order parameter
\footnote{For simplicity, we neglected the local-field corrections, which have similar effect on dipoling and octupoling, and are not important for the order-of-magnitude estimations.}.
Therefore, we propose a revised criterion for octupoling temperature in the very weak poling regime
	\begin{align}
	T_\text{v.w.} = \alpha \frac{\DE}{\kB},
	\label{eq:VeryWeakPolingCriterion}
	\end{align}
instead of the weak criterion $T_{\rm weak} \approx {\DE}/{\kB}$ from Ref.~\onlinecite{MitusPZ1}. The value of the octupolar order parameter in the very weak poling regime is obtained by evaluating Eq.\ \eqref{eq:avg-ord-par} at $(\beta^\star)_\text{v.w.}=\frac{1}{2\alpha}$ (compare Eqs.\ \eqref{eq:beta*} and \eqref{eq:VeryWeakPolingCriterion}):
\begin{equation}
\label{eq:very-weak-001}
|P|_\text{v.w.} \approx 10^{-2},
\end{equation}
in fair agreement with its dipolar counterpart, Eq.\ \eqref{eq:two-state}.
	
\begin{figure}
		\centering
		\includegraphics[width=\mycolumnwidth]{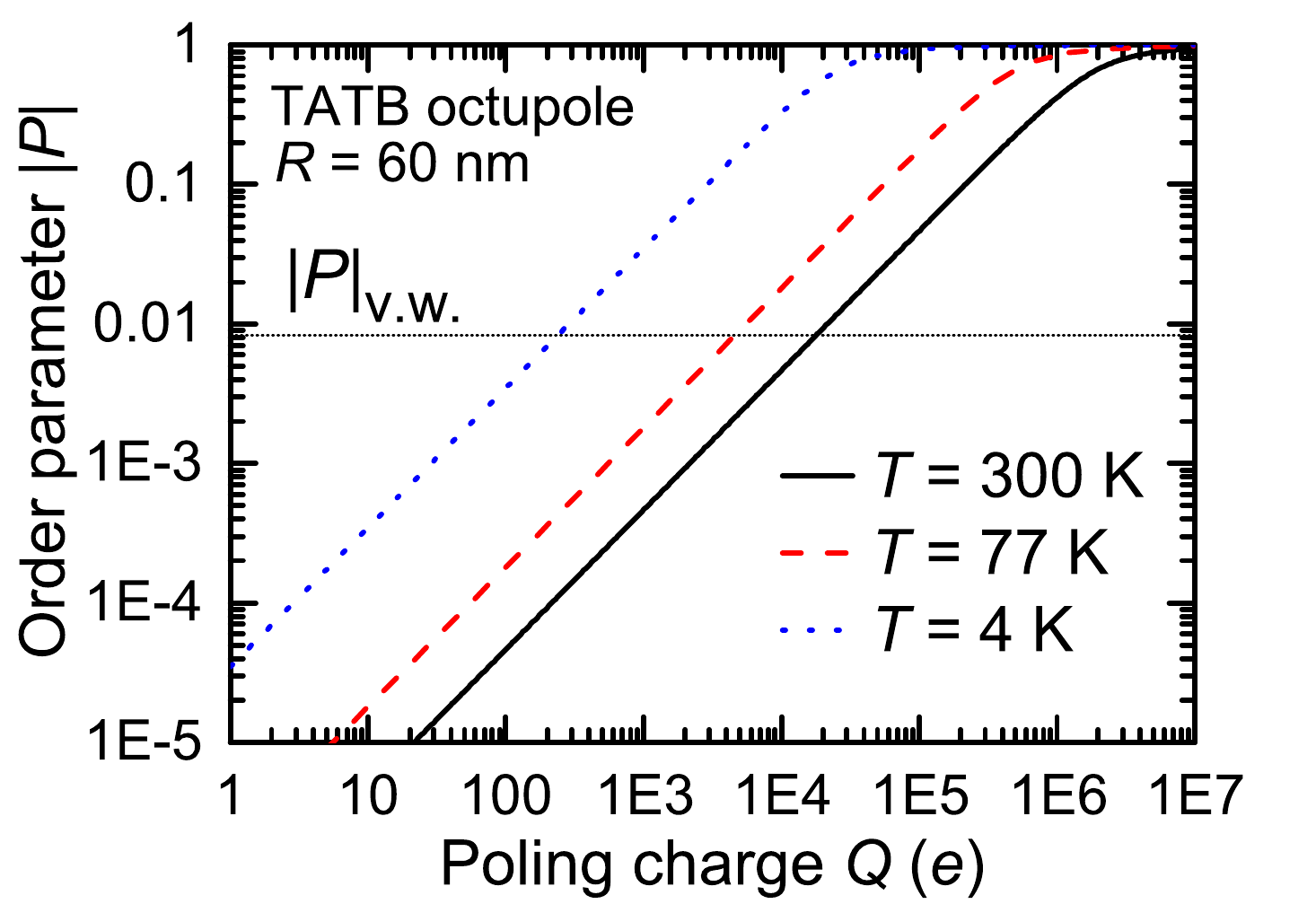}
		\caption{Order parameter $|P|$ in the central part of a point charge poling cell in function of the poling charge $Q$ for selected temperatures. The  order parameter $|P|_\text{v.w.}$ for the very weak regime is represented by the horizontal line.
		}
		\label{fig:PofQforT}
\end{figure}

\subsubsection{Discussion of octupoling conditions}
\label{s:polingconditions}
	
Based on previous considerations, we discuss now the conditions for effective octupoling in the central region of a realistic poling cell, in the very weak poling regime, Eq.\ \eqref{eq:very-weak-001}.	In order to increase the poling temperature
the cell-related factor $Q/R^4$, see Eq.\ \eqref{eq:barrier6QOct}, has to
be maximized;  therefore we look for small values of $R$ and large values of $Q$. In what follows we use the same poling cell size $R=60$~nm as in Refs.~\onlinecite{MitusPZ1,ZyssAMARIS}, and concentrate on the analysis of physical constraints imposed by the magnitude of the poling charge $Q$.
	
Figure \ref{fig:PofQforT} shows the double logarithmic plot of order parameter  $|P|=I_{1,0}(\beta^\star(Q,T))$ (Eq.\ \eqref{eq:avg-ord-par}) vs. $Q$  for a few selected temperatures - liquid Helium, liquid Nitrogen and room temperature. It implies the power law dependence for $|P|$:
\begin{equation}
\label{eq:P_power_law}
|P|(Q, T) = f(T)\,Q^{x}.
\end{equation}
The exponent $x$ and  function $f$ can be easily inferred in the high temperature expansion (HTE)  limit when $\beta^\star \to 0$. Namely, in this limit $I_{1,0}(\beta^\star) \approx \frac{1}{2} \beta^\star \propto \frac{Q}{T}$. Hence, $x=1$ and $f(T) \propto 1/T$. A closer inspection of the plot shows that those results hold for, say, $|P| \leq 0.5$, far beyond the very weak order regime. Putting $|P| = |P|_\text{v.w.}$ in this formula  and using the definition of $\beta^{\star}$ in Eq.\ \eqref{eq:beta*} provides the charge $Q_\text{v.w.}$ which grants the very weak poling conditions at temperature $T$:
\begin{equation}
\label{eq:Q_T_vw}
Q_\text{v.w.} = \frac{4}{15}\,\frac{k_B R^4}{k_0 \norm{\mathcal{O}}}\, |P|_\text{v.w.}\,T.
\end{equation}

The practical implementation of very weak octupolar poling conditions at temperature $T$ is limited by the restrictions imposed on the cell setup by  the acceptable magnitude of the accompanying electric field ${\cal E}(Q_\text{v.w.})$. Two physical effects appear to be of primary importance.

Firstly, the dielectric strength of the medium must prevent electric breakdown. In practice, the ``point'' poling charges are small charged spheres of radius $r_s$. The electric field $\mathcal{E}_s$  close to the surface of a sphere bearing charge $Q$ is
	\begin{equation}
	\mathcal{E}_s(Q)=\frac{k_0 Q}{{r_s}^2}. \label{eq:6QrequiredEs}
	\end{equation}
The dielectric strength of the matrix has to be higher than $\mathcal{E}_s(Q)$, which sets an upper boundary $Q_{s,max}$ on $Q$: $Q < Q_{s,max}$.

Secondly,  Coulomb forces between poling charges must be balanced by elastic reaction forces, so as to prevent collapse of the poling cell. The net Coulomb force acting on any of the poling charges due to the remaining five ones (see Fig.~\ref{fig:poling-cell-charges}) is directed towards the center of the poling cell and is given by
	\begin{equation}
	F(Q) = c_F \frac{k_0 Q^2}{R^2} , \; c_F=\frac{15-4\sqrt{3}}{12}   \approx 0.67. 
	\label{eq:6QForce}
	\end{equation}
The charged spheres of radius $r_s$   interact through a surface $S\approx\pi {r_s}^2$ with the surrounding medium. The resulting pressure is:
	  \begin{equation}
	   \eta(Q)=\frac{F}{S}=c_F \frac{k_0 Q^2}{\pi {r_s}^2 R^2}.
	   \label{eq:pressure}
	  \end{equation}
To avoid the mechanical breakdown of the polymer matrix its tensile strength has to be larger than $\eta(Q)$, which sets another upper limit $Q_{\eta, max}$ on the magnitude of $Q$: $Q < Q_{\eta, max}$.

\begin{figure}
		\centering
		\includegraphics[width=\columnwidth]{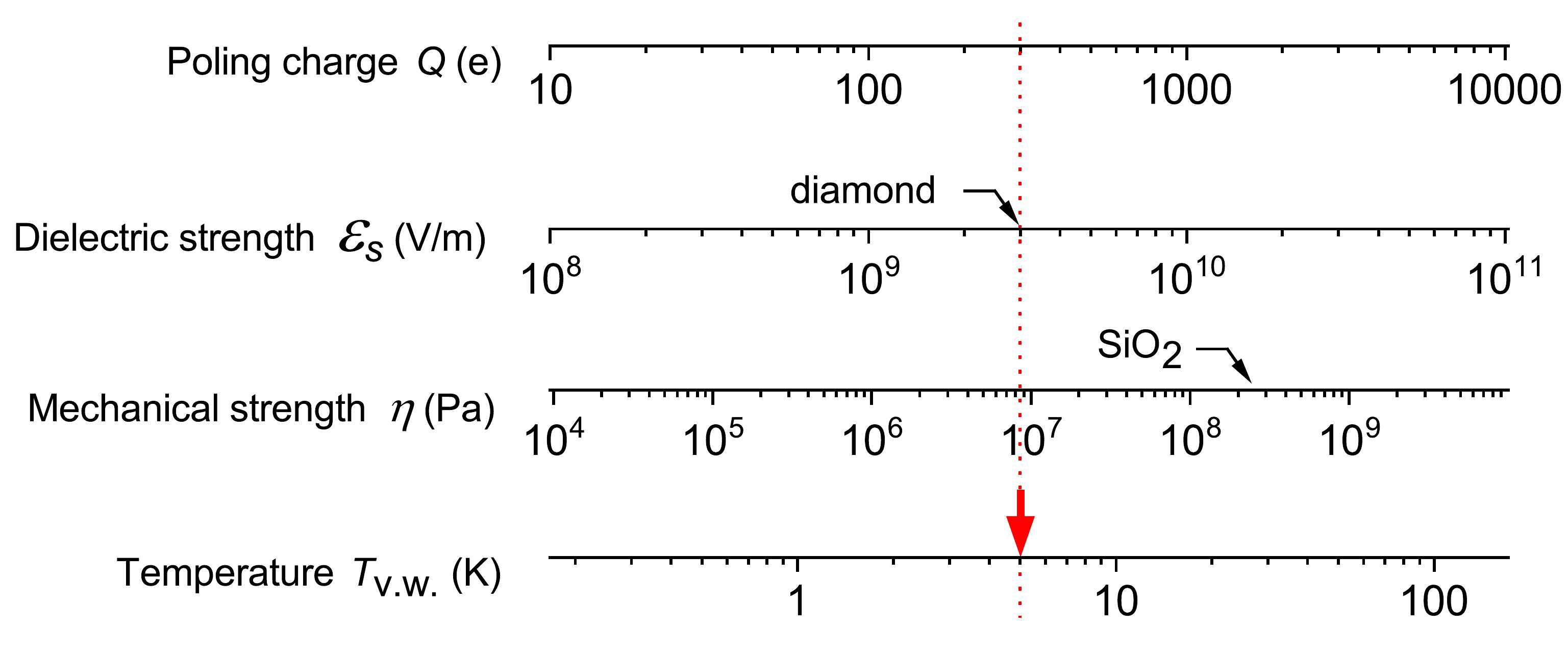}
		\caption{
Mutual relations between the poling charge $Q$, the dielectric strength $\mathcal{E}_s(Q)$ (Eq.\ \eqref{eq:6QrequiredEs}) and mechanical strength $\eta(Q)$ (Eq.\ \eqref{eq:pressure}), and the temperature $T(Q)$ (Eq.\ \eqref{eq:Q_T_vw}) for octupoling at the cell's center in the very weak poling regime for $R=60$ nm and $r_s=12$ nm. Black arrows design maximal accessible  values (dielectric and mechanical strengths) of the poling setup. The red arrow points at maximal accessible temperature $T_\text{v.w.}$ resulting from the restrictions imposed by $\mathcal{E}_s$.
		}
		\label{fig:pointChargesBarrierForce}
\end{figure}

The mutual relations between the poling charge $Q$,  the electric field amplitude $\mathcal{E}_s(Q)$ (Eq.\ \eqref{eq:6QrequiredEs}), the pressure $\eta(Q)$ (Eq.\ \eqref{eq:pressure}) and temperature $T(Q)$ (Eq.\ \eqref{eq:Q_T_vw})  are summarized in  Fig.~\ref{fig:pointChargesBarrierForce}. They are useful for an evaluation of very weak poling conditions at given temperature $T$. To this end, the parameters $Q_{s,max}$ and $Q_{\eta,max}$ have to be estimated.

Let us discuss $Q_{s,max}$ first. The  dielectric strength of common polymers is at least one order of magnitude higher than the dielectric strength of air.\cite{GunterNLOMaterialsBook, PolymerDielectricStrength} Moreover, it usually increases for small thicknesses of the insulating layer, short timescale, low temperature and for high purity materials.\cite{PolymerBook}	 Interestingly, thin films (170 nm) of thermally-cured DNA--CTMA sol--gel have been demonstrated \cite{GroteDNACTMADielectricStrength} to sustain an electric field as high as $9\cdot10^8\frac{\rm V}{\rm m}$  at room temperature. Moreover, the spheres can be covered with an insulating layer to increase the threshold for electric breakdown.  An outstanding value of dielectric strength is the avalanche breakdown strength of chemical vapor deposited diamond.\cite{Diamond1983SovPhys,DiamondElectronicMat,DiamondLuminescence5e8V-m,Diamond-3e9} It has been reported \cite{Diamond-3e9} that an electric field up to $5 \cdot 10^9~\frac{\rm V}{\rm m}$ can be applied to a 200 nm-thick diamond.
To estimate the threshold parameter $Q_\text{s,max}$ we set $\mathcal{E}_s$ to a slightly lower limit of $3 \cdot 10^9~\frac{\rm V}{\rm m}$ (at the onset of breakdown \cite{Diamond-3e9}) and use the value   $r_s=R/5$ for the radius of the sphere. We find, from Eq.\ \eqref{eq:6QrequiredEs},  $Q_{s,max} \approx 300\,e$.

Next, let us estimate the mechanical breakdown effect related to pressure.
Tensile strength of PMMA    is about 50-75 MPa at room temperature.\cite{PMMA-tensileStrength} The corresponding upper limits for the poling  charge $Q_{\eta,max}$ are correspondingly in the $700~e$ to $900~e$ range.
Moreover, the charged spheres can be mechanically supported by a skeleton made from a harder material, \eg fused silica ($\text{SiO}_2$) with tensile strength of about 150 MPa.\cite{DielectricStrengthSiO2} Then, $Q_{\eta,max}$ can raise up to $1250\,e$. 	

To summarize, the critical restrictions on $Q$ result from the electric breakdown effect:
\begin{equation}
\label{eq:Q_max}
Q < Q_{s,max} \approx 300\,\mathrm{e} < Q_{\eta,max}.
\end{equation}
A gold sphere of radius $r_s= R/5=12$~nm contains about $4\cdot10^5$ atoms and $4\cdot10^3$ laying on its surface, so charging it with 300 electrons results in a realistic charge density.

Finally, inserting $Q_\text{v.w.} = Q_{s,max} = 300$~e and $|P|_\text{v.w.} = 10^{-2}$ into Eq.\ \eqref{eq:Q_T_vw} we find that the very weak octupoling can take place in the liquid Helium range:
	\begin{equation}
	T_\text{v.w.}\approx5~K.
	\label{eq:Temp5K}
	\end{equation}
	
\subsection{Dipoling-induced octupolar order}
\label{sect:dipoling}
	
A  different way of promoting octupolar orientational order is to consider molecules of mixed dipolar--octupolar character, to be poled by a strong dipolar electric field. 

By way of introducing a quantitative model, let us start with a qualitative presentation.
It is well known that symmetry arguments strictly forbid 
octupoling of dipolar--octupolar molecules via a homogeneous external field in the weak poling regime, \emph{i.e.}, when the Boltzmann factor is expanded to the first order w.r.t.\ the inverse temperature $\beta$. \cite{BZ98,ZyssAMARIS}
However, higher order expansion terms may exhibit the symmetry features of octupolar order parameter and therefore promote a non-zero octupolar order.
For a mixed dipolar--octupolar molecule sustaining a strong octupolar second polarizability tensor component,  SHG anisotropy may exhibit properties with octupolar symmetry features. A  more technical development of this situation is given in the next Section.

More generally, we wish to point out that such a mechanism complies with the basic distinction between low- and high-field effects that pervades throughout nonlinear optical phenomenology. 
In a similar context it has been shown both theoretically and experimentally that the template mixed dipolar--octupolar molecule  
1,3-dinitro-4,6-di-(n-butylamino)-benzene (DNDAB,  with $C_{2v}$ point-group symmetry, see Ref.\ \onlinecite{BZ98})  exhibits a significant departure of its nonlinear anisotropy from the weak poling field value of 3, onto lower values that are indicative of strong octupolar contributions to nonlinear susceptibility.\cite{DNDABreport,IreneCazenobePhD} In addition to the usual dipolar component, SHG signal has been shown to sustain octupolar properties.  

Let us formulate this scenario in a quantitative way.
Consider  a planar dipolar--octupolar molecule abiding to an in-plane two-fold symmetry axis and a dipole moment along that axis  (\ie in $C_{2v}$ symmetry, lowering the symmetry of an equilateral triangle into that of an isosceles one), in an external homogeneous field $\vec{\mathcal{E}}$. Generalization to the case where the poling field exhibit an octupolar component is briefly discussed in Sect. \ref{sect:discussion}.
The energy $E(\varphi)$  of an octupolar molecule with attached dipole moment $\vec{\mu}$ in this case originates exclusively from the dipole--field interaction, since the field $\vec{\mathcal{E}}$ is not allowed to couple, due to symmetry arguments, to the octupolar electric moment of the molecule.\cite{BZ98,ZyssAMARIS} $E(\varphi)$  depends on the orientation $\varphi$ of the dipole (note different energy formula from that in Eq.\ \eqref{eq:avg-ord-par}):
\begin{equation}
E(\varphi)=-\vec{\mathcal{E}}\cdot\vec{\mu}=-\mathcal{E}\mu\cos\varphi.
\end{equation}

\begin{figure}
	\centering
	\newdimen\w
	\w=1.0\mycolumnwidth
	\includegraphics[width=\w]{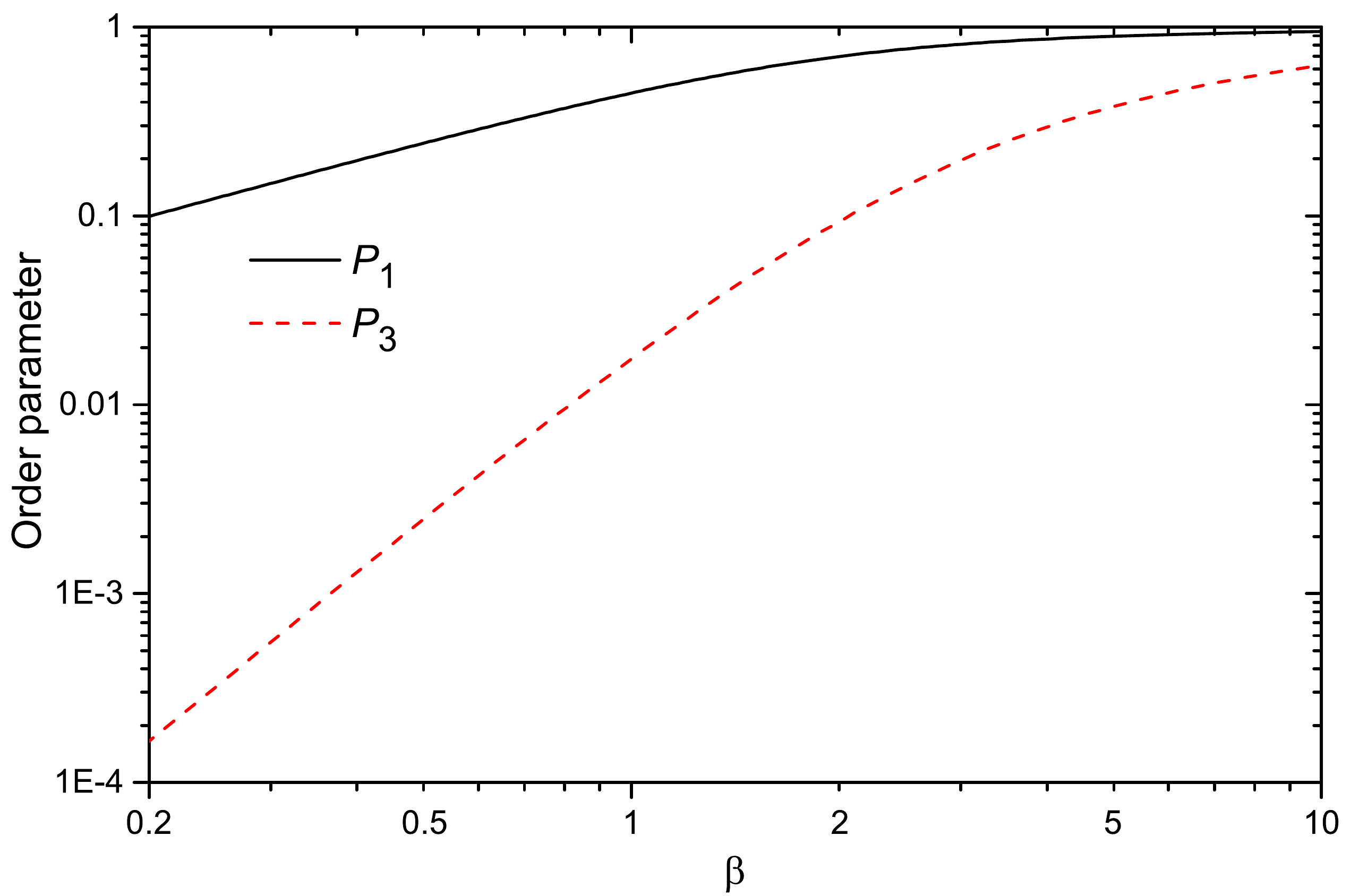}%
	\adnote{0.605\w}{0.097\w}{\includegraphics[width=0.575\w]{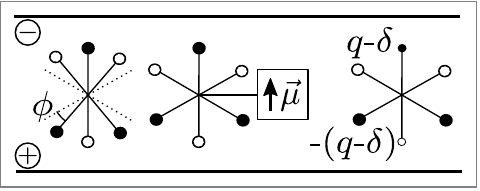}}%
	\caption{Log-log plot of octupolar order parameters $P_1$ (black solid line) and $P_3$ (red dashed line) in function of the inverse temperature $\beta^*_d$. Inset: modified dipolar--octupolar molecules, see text.}
	\label{fig:p3-vs-beta}
\end{figure}

\begin{figure}
        \centering
		\newdimen\w
		\w=\mycolumnwidth
		\includegraphics[width=\w]{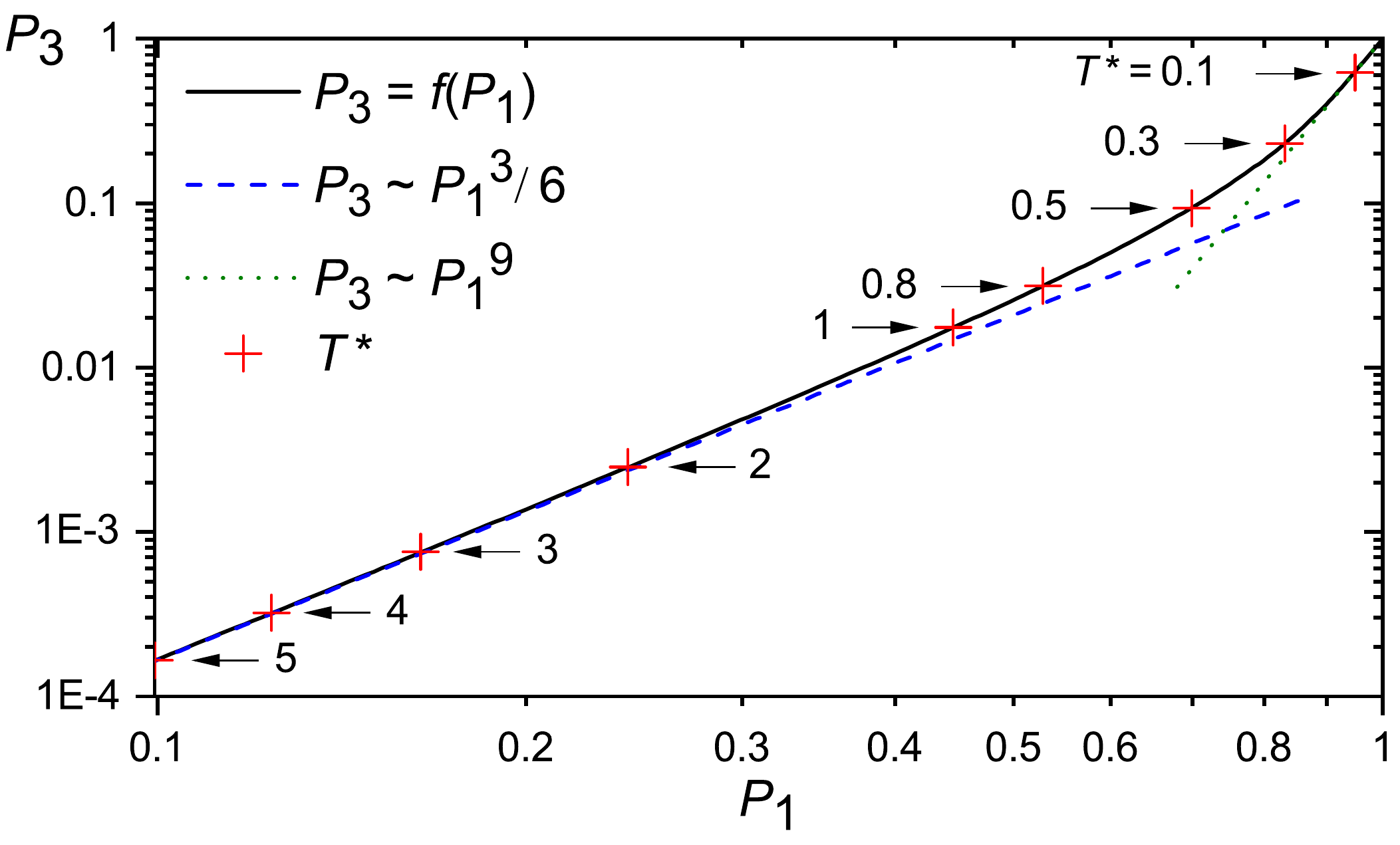}%
		\caption{Log-log plot of parametric relation between order parameters
        $P_3$ and $P_1$ (black solid line). The reduced temperature parameter $T^\star=\frac{1}{\beta^\star_d}$  is marked along the curve.
		The high-temperature asymptote $P_3=(P_1)^3/6$ (blue dashed line)
		and low-temperature asymptote $P_3=(P_1)^9$ (green dotted line) are
		 also shown. }
		\label{fig:p3-vs-p1}
\end{figure}

The average local order planar parameters $P_n$, Eq.\ \eqref{eq:Pav},  read:
\begin{equation}\label{eq:dipoling-ord-par-exact}
\begin{split}
P_n= & \langle e^{i n \varphi}\rangle = \frac{1}{Z} \int_0^{2 \pi} e^{i n \varphi} e^{- \beta\,E(\varphi)} {\,\rm d}\varphi \\
& = \frac{1}{Z}\int_0^{2 \pi} \cos(n \varphi) e^{ \beta^\star_d \cos \varphi} {\,\rm d}\varphi = \frac{I_n(\beta^\star_d)}{I_0(\beta^\star_d)},
\end{split}
\end{equation}
where $Z = \int_0^{2 \pi} e^{ \beta^\star_d \cos \varphi} {\,\rm d}\varphi$ and
$\beta^\star_d =\frac{\mathcal{E}\mu}{k_B T}$. The plots of dipolar and octupolar order parameters $P_1$  and $P_3$ in function of $\beta^\star_d$ are shown in Fig.~\ref{fig:p3-vs-beta}. At low temperatures  $P_3 \approx (P_1)^9$ (low temperature expansion) and both parameters have comparable values when $P_1$ is close to 1. The situation becomes very different at high temperatures
(weak interaction limit) --  the leading terms in this expansion read
\begin{equation} \label{eq:Pn_vs_beta}
P_n\simeq\frac{(\beta^\star_d)^n}{2^n n!}.
\end{equation}
In particular, the octupolar order parameter $P_3$ vanishes in the first order of high-temperature expansion, in agreement with our former statement. However, it becomes non-zero for the third order expansion, while relating to the dipolar order parameter $P_1$ in the following way:
\begin{equation}
	P_3\simeq(P_1)^3/6,
\end{equation}
indicating that the octupolar order parameter becomes negligibly small in comparison with the dipolar one when $\beta^\star_d \ll 1$.
On the other hand, a sufficiently strong homogeneous field (such that $P_1 \approx 1$)  promotes a high degree of  in-plane octupolar order. 
The relation between  order parameters $P_3$ and $P_1$ for a range of temperatures is summarized in Fig.~\ref{fig:p3-vs-p1}.

Let us estimate the magnitude of the dipolar moment necessary  to support the
very weak poling regime $(P_3)_{\text{v.w.}} = 0.01$, Eq.\ \eqref{eq:very-weak-001}, at Nitrogen temperatures ($T=77$ K). Using  Eq.\ \eqref{eq:Pn_vs_beta} and the definition of $\beta^\star_d$ we find
\begin{equation} \label{eq:TvW_vs_mu}
T \simeq \frac{\mathcal{E}\mu}{k_B \sqrt[3]{48 (P_3)_{\text{v.w.}}}} \approx 1.28 \frac{\mathcal{E}\mu}{k_B} \approx 61.7\, \tfrac{\rm K}{\rm D}\, \mu,
\end{equation}
with $\mathcal{E}=2\cdot10^8$~V/m as available poling field strength;\cite{GunterNLOMaterialsBook} $\mu$ is expressed in Debye unit. Upon replacing $T$ with $77$ K we get $\mu \simeq 1.25$ D. The related dipolar order parameter $P_1$ is, contrary to $P_3$, not negligible: $P_1 \approx 0.36$. To create such a dipole the 6AO molecule has to be modified accordingly.  Let us discuss briefly three strategies: (i) distorting the shape of the octupolar molecule, (ii) adding a peripheral dipolar group, and (iii) modifying the molecular charge density (see inset in Fig.~\ref{fig:p3-vs-beta}).
	
In the first case  four arms of a 6AO molecule are bent towards its $C_2$ axis (formed by remaining two arms) by $\phi=2.7^\circ$. Such a distortion removes the $C_3$ symmetry axis but leaves the $C_2$ unchanged (\ie lowers symmetry from $D_{3h}$ to $C_{2v}$) and decreases the irreducible octupole moment by about 1\% while inducing a dipole moment of 1.25 D.

In the second strategy, a group with a small dipole moment is rigidly attached to an octupolar molecule in such a manner that it is separated from the $\pi$-conjugated system responsible for octupolar hyperpolarizability (in a similar way to the CN nitrile group in the NPAN molecule \cite{BZ98}). For example, it  could be accomplished by a substitution of one C--H bond by a more polarized C--Cl bond. To estimate the magnitude of the corresponding dipole moment, we have analyzed experimental dipole moment data \cite{CRCDipole} for 18 simple compounds consisting of carbon, hydrogen, and a single chloride atom adjacent to a carbon atom, such that with H replacing Cl their dipole moment would vanish due to symmetry. The average and standard deviation for those compounds is $\mu_\text{CCl}=(1.95\pm0.23)$~D. Similar query for 14 compounds containing a single CN group yields $\mu_\text{CN}=(3.96\pm0.30)$~D.

In the third template the magnitude of two $\pm q$ opposite charges  in the 6AO molecule is reduced by  $\delta=0.08 \, q$. It leads again to $C_{2v}$ symmetry and decreases the irreducible octupole moment by 2.5\% while inducing a dipole moment of 1.25 D.   This method  is inspired by the generalized equivalent internal potential model.\cite{YaronOproptoB}  Namely, in the TATB molecule one pair of $\mathrm{NO}_2$ and $\mathrm{NH}_2$ donor and acceptor groups (which are strong substituents in nonlinear regime\cite{YaronOproptoB}) are exchanged with other two slightly less ``pushing'' and ``pulling'' substituents.

We point out that the modification of the structure of  purely octupolar molecule is, in general, accompanied by a change of the octupolar component of the molecule's second hyperpolarizability. On the other hand, taking into account the fact that the structural modifications as well as the changes of irreducible octupolar moments  are minor, we assume that the octupolar non-linear responses are also modified to a low degree. A short discussion of this topic is presented in Sect.~\ref{sect:discussion}.

We conclude that the very weak octupolar order can be reached by di-poling of dipolar--octupolar molecules with $\mu=1.25$~D at liquid Nitrogen temperatures:
\begin{equation}
T_\text{v.w.} = 77 \text{ K}.
\end{equation}
	
\section{SHG coefficients for different scenarios for promoting nano-octupolar order}	\label{s:NLO-properties}
	
Let us estimate the values of some experimentally-accessible NLO parameters  in the previously discussed poling scenarios. To this end, we apply the methodology worked out in Ref.~\onlinecite{ZyssAMARIS}, aiming at the calculation of NLO susceptibilities of a system of  multipolar molecules ordered by multipolar electric field	in the weak interaction (HTE) limit. Namely, the  average of any tensor property attached  to a molecule (e.g. its charge distribution) can be expressed as a weighted sum of irreducible tensorial parts of the property, weighed by multipolar local order parameters $P_n = \langle p_n \rangle$ (Eq.\ \eqref{eq:Pav}). For example, the quadratic nonlinear susceptibility tensor $\tilde{\chi}^{(2)}$, defined as the product of number density $\mathcal{N}$ and quadratic hyperpolarizability tensor $\tilde{\beta}$ averaged over molecular orientations, reads in 2D \cite{ZyssAMARIS}
\begin{equation}\label{eq:JZchi2avg}
 \tilde{\chi}^{(2)}(\vec{r})=\mathcal{N}\langle\tilde{\beta}\rangle=\mathcal{N}\, \sum_{n=0}^{3} P_n(\vec r) \,\tilde{\beta}^{J=n},
	\end{equation}
where  tildes denote tensors.  The sum runs over the irreducible parts $J$ of the tensorial property $\tilde{\beta}$  -- from $J=0$  (scalar part) to the rank of the tensorial property $J=3$ for $\tilde{\chi}^{(2)}$. Moreover, a symmetric tensor, like $\tilde{\beta}$ in Kleinmann regime, consists only of irreducible parts with  $J$ displaying the same parity as the rank of the tensor.\cite{Jerphagnon} Thus,  the NLO response consists of vectorial ($J=1$) and octupolar ($J=3$) parts:
\begin{equation} \label{eq:JZchi2avg-1-3}
\tilde{\chi}^{(2)}(\vec{r})=\mathcal{N}\,\big(P_1(\vec r) \,\tilde{\beta}^{J=1} +  P_3(\vec r) \,\tilde{\beta}^{J=3}\big).
\end{equation}

For a three-fold-symmetric octupolar molecule like TATB,   the vectorial part $\tilde{\beta}^{J=1}$ vanishes. In the case of dipolar induced octupolar order (Sect.\ \ref{sect:dipoling}) the vectorial part becomes, in general, non-zero, and can dominate the response as $P_3 \ll P_1$ in the weak interaction limit, see Fig.~\ref{fig:p3-vs-beta}.
However,  the octupolar part of the response offers advantages over its vectorial counterpart due to its richer tensorial structure.\cite{ZyssJCP93engineeringImplications,BZ98} In particular, while the octupolar response in a given (vectorial part) direction can be dominated by its dipolar counterpart, it remains uninfluenced  in two other directions. In other words, the optical waves in different directions can still be nonlinearly coupled by non-zero off-diagonal tensor components like $\chi^{(2)}_{xyy}, \chi^{(2)}_{yxx}$. In what follows, we focus on octupolar order and proceed to estimate the value of
\begin{equation} \label{eq:JZchi2avg-3}
\tilde{\chi}^{(2)}(\vec{r})=\mathcal{N}\,  P_3(\vec r) \,\tilde{\beta}^{J=3}.
\end{equation}

Using the overall global order parameter $\mathbb{P}$, Eq.\ \eqref{eq:Pavglobal}, instead of $P_3$ in Eqs.\ \eqref{eq:JZchi2avg}--\eqref{eq:JZchi2avg-3} corresponds  to averaging over the positions $\vec{r}$ in the cell, and yields the global susceptibility of the system $\tilde{\chi}^{(2)}$. A related, more widely used parameter, is the nonlinear susceptibility coefficient matrix \cite{BoydNLO}	 $d_{il}=\tfrac{1}{2\varepsilon_0}\chi^{(2)}_{ijk}$ ($\varepsilon_0$ is vacuum permittivity, Voigt's index notation \cite{BoydNLO} is used), for which we get
	\begin{equation}\label{eq:d_il-final}
    	d_{il}=\frac{\mathcal{N}}{2\varepsilon_0}\, \mathbb{P} \, \beta^{J=3}_{ijk}.
	\end{equation}

Eqs.\ \eqref{eq:JZchi2avg} -- \eqref{eq:d_il-final}, which were derived within the weak interaction limit, can as well be applied  for order parameters obtained by other means,  provided that the weak interaction approximation remains valid -- in particular, in the very weak poling limit used in this paper.

Let us estimate the order of magnitude of the $d_{il}$ tensor components for  TATB guest molecules homogeneously dispersed  in a PMMA polymer host matrix with weight fraction $f_{\rm wt}$.
The number density  is $\mathcal{N}=f_{\rm wt} \rho / m_1$, where $\rho$ denotes the density of the system and $m_1$ stands for the mass of one molecule.
It is useful to estimate the corresponding average intermolecular distance $l$ between guest molecules. For uniformly mixed system  $l \approx\mathcal{N}^{-1/3}$, and the reduced distance $L = l/d$ reads
	\begin{equation}
	L=(f_{\rm wt} \rho / m_1)^{-1/3} d^{-1} \approx 2.92 / \sqrt[3]{f_{\rm wt}},
	\label{eq:L-vs-f}
	\end{equation}
with $m_1=258$~u (atomic mass unit), $d=2.44$~\AA\ for a TATB molecule and
$\rho=1.18~\text{g}/\text{cm}^3$ for pure PMMA polymer matrix. Next, the components of the $\beta^{J=3}_{ijk}$ hyperpolarizability of a TATB molecule in the reference frame of Fig.~\ref{fig:TATB-xy} are estimated as \cite{JACS1992,ZyssJCP93engineeringImplications}
$\beta_{xxx}=-\beta_{xyy}=10^{-29}\text{ [esu]}\approx3.5\cdot10^{-50}\,\tfrac{(C\,m)^3}{J^2}\text{ [SI]}$.

The  results are summarized in Table~\ref{tab:NLOsuscept}, where we use three values of intermolecular distance $L = 5, 10$ and $15$. The first one corresponds to $f_{\rm wt}\approx 0.2$
which is a high but achievable concentration in host--guest poled polymer systems.\cite{GunterNLOMaterialsBook} A short comment on this choice is given in Sect.\ \ref{sect:discussion}. Two simplifications were made. Firstly, in the case of point charge octupoling the order is inhomogeneous and, for simplicity, only the central region of the cell is considered, where $\mathbb{P}\approx P(r=0)$, see Fig.~\ref{fig:6QGS-orientation}. Secondly,
in the case of dipoling-assisted scenario we have neglected the influence of added dipolar moment on  the hyperpolarizability $\beta_{ijk}$, see comment at the end of the previous Section.
	
	\begin{table}[h]
		\centering
		\begin{tabular}{ccccc}
			\hline\hline
			  L& & $f_{wt}$ & ${\cal N}$ [cm$^{-3}$] & $d_{11}$ [pm/V] \\
			\hline
			 5 & & 20\% & $5 \cdot 10^{20}$ & $1.5\cdot10^{-2}$ \\
			 10 & & 2.5\% & $6 \cdot 10^{19}$  & $1.9\cdot10^{-3}$ \\
			 15 & & 0.7\% & $2 \cdot 10^{19}$ & $5.6\cdot10^{-4}$ \\
			\hline\hline
		\end{tabular}
		\caption{NLO susceptibility tensor  $\tilde{d}=(2\varepsilon_0)^{-1}\tilde{\chi}^{(2)}$ in the very weak poling regime, Eq.\ \eqref{eq:d_il-final}, for a few values of relative distance $L$, weight fraction $f_{wt}$ and concentration ${\cal N}$.
		The non-zero coefficients of $d_{il}$ are $d_{11}=-d_{21}=-d_{26}$.
		}
		\label{tab:NLOsuscept}
	\end{table}

For comparison, coefficients  $d_{il}$ for NLO crystals are typically of the order of 0.5 pm/V (quartz) to 70 pm/V (GaSe). For dipoled host--guest system in polymer matrix the achievable values are 2.5 pm/V (DR1, 2.74 wt\% in PMMA) to 84 pm/V (modified DR1, 10 wt\% in PMMA).\cite{GunterNLOMaterialsBook,JeongOctupolarMoleculesNonlinear2015} In the next Section we discuss briefly some scenarios which result in a substantial increase of the value of this parameter in our model.

\section{Summary and discussion}
\label{sect:discussion}

We have studied two scenarios of orientational ordering of small flat octupolar and slightly modified octupolar molecules by external electric field in the context of quadratic non-linear optics phenomena with emphasis on second harmonic generation. While we have used a simple statistical mechanics modeling, a new methodological approach was applied to the study of the very weak poling regime, when the emerging overall  orientational order is much smaller than thermal fluctuations. Experiments should be enabled by
the improved sensitivity and higher signal-to noise ratio of current detectors used in NLO experiments.

Our method of analysis of very low orientational octupolar order encompasses  both statistical mechanics modeling and an estimate of material limitations resulting from current nanotechnologies, with detailed calculations performed for the TATB molecule. However,
the method is very general and can be applied to larger molecules (see below), metal-organic complexes, nanocrystals, nanostructures etc.

We have found that the current nano-scale technology limits the octupoling temperatures for small octupolar molecules, like TATB, to a few Kelvins. This improves the previous estimations \cite{MitusPZ1} by four orders of magnitude.  Its origin is as follows: (i)  increase of the poling temperature by  factor $\alpha\approx30$ due to the very weak poling scheme, (ii) five times larger octupole moment and (iii) much stronger octupoling electric field. Namely, the electric field strength in the middle of the poling cell would correspond to the voltage $ V_0 = \frac{15\pi}{16}\frac{ k_0\, Q }{R} \approx 35\text{ V}$ in the electrode poling scenario,\cite{MitusPZ1} where the value $V_0=0.1$~V has been previously used. The combination of those numbers yields around $5\cdot10^4$ -- a factor by which the poling temperature was underestimated in Ref.~\onlinecite{MitusPZ1}.

In the case of dipoling --  a scenario in which modified octupoles with a small dipolar moment are poled by a strong homogeneous electric field -- the very weak octupolar order is preserved  up to liquid Nitrogen temperatures. The characteristic feature of this scenario  is a trade-off between an increase of the poling temperature, proportional to $\mu$, Eq.\ \eqref{eq:TvW_vs_mu}, and the loss of the purely octupolar component of the NLO response. The numerical estimations were based on the assumption that the emergence of a small dipolar moment does not influence the  octupolar hyperpolarizability.  This hypothesis is based on the model \cite{YaronOproptoB} of equivalent internal potentials acting on the polarizable molecular skeleton from the surrounding push--pull substituent sites. Both  permanent moments and corresponding irreducible parts of the hyperpolarizability are linear responses to the internal potentials, and are therefore expected to change in the same proportion for a small perturbation of the molecular structure.
In particular, a correlation between calculated $(J=3, m=\pm 3)$ components of $\tilde{\mathcal{O}}$ and $\tilde{\beta}$ has been reported.\cite{YaronOproptoB} A quantitative analysis of this topic requires quantum chemistry  calculations for the molecules of interest and goes beyond the scope of this paper. Another advantage of this scenario is a strong  increase of octupolar order at lower temperatures. At Helium temperatures ($T=5$K) the order parameters are $P_1 \approx 0.96, P_3 \approx 0.68$.  The latter indicates a high degree of octupolar order, some two orders of magnitude larger than in the very weak poling limit. This yields, in the first approximation, an increase of SHG by an order of magnitude: $d_{11} \approx 0.1$ pm/V for $f_{wt} = 20\%$, see Table \ref{tab:NLOsuscept}.
		
The results of this study imply that a moderate change of the parameters of the model can raise the octupoling temperature to a more acceptable range. We postpone a systematical analysis to future studies and will limit ourselves here to the discussion of some emerging scenarios. Consider first the pure octupoling method. Technological constraints impose an upper limit to the poling charge $Q_{s, max}$ (Eq.\ \eqref{eq:Q_max}). This introduces (see Eq.\ \eqref{eq:Q_T_vw})
a relation between the poling temperature, the octupolar order parameter and the size of the poling cell in the very weak poling regime:
\begin{equation}
T = \text{const}\,\frac{\norm{\mathcal{O}}}{R^4\, |P|_\text{v.w.}} \propto \frac{|q|\,d^3}{R^4\, |P|_\text{v.w.}},
\end{equation}
since $\mathcal{O}\propto qd^3$. We find that the most promising method to increase the poling temperature is to increase the octupolar moment of the molecule. For example, a twofold increase of charge $q$ accompanied by a twofold increase of $d$ (thus the size of a molecule) brings the octupoling temperature to the Nitrogen range; if the size were to be increased by a factor of three, the poling temperature would still be below room temperature but close to it. The corresponding molecular design provides an interesting target for quantum chemistry calculations. Other than that, reduction of the  detection limit of octupolar order $P$ below the $P\approx1\%$  threshold would also allow for an increase of the poling temperature.

Another interesting scenario not discussed in this paper is the dipole-assisted octupoling, i.e., in the presence of both dipolar and octupolar poling fields which jointly enhance the octupolar order. When both octupolar-order components (due to dipolar and octupolar fields) are in the very weak poling regimes, then the resulting octupolar order is the sum of both components. This is, e.g., the case for larger molecules discussed above, with octupoling temperature in Nitrogen region, the net order remaining in the very weak limit, leading to a very low NLO response. On the other hand, when both components are not small then the net octupolar order is larger than the sum of both components, due to nonlinear interaction term. Consider again the  larger molecule discussed above (with octupole moment increased by factor 16 and dipole moment $\mu = 1.25$ D) now at Helium temperature ($T=5$ K). Then, the net octupolar order parameter $P_3$ has two components: the dipoling-induced octupolar order parameter $P_3 \approx 0.68$ and the octupoling-induced one $P_3 \approx 0.13$. The sum yields around 0.8, but the nonlinear effects will still increase it, approaching the limit of perfect octupolar order $P_3 = 1$.
In this limit the nonlinear susceptibility tensor $\tilde{\chi}^{(2)}$
(Eq.\ \eqref{eq:JZchi2avg} is directly proportional to the quadratic hyperpolarizability tensor $\tilde{\beta}$. For TATB molecules we find $d_{11} \approx 1$ pm/V  for $f_{wt} = 20\%$. Detailed analysis of this topic goes beyond the scope of this paper.

The next step in the modeling of octupoling  is to account for the three-dimensional geometry of more general molecular octupoles. The description of orientational order in three dimensions requires a more advanced mathematical formalism. This extension is currently under investigation and will be reported later.

Current study did not account for intermolecular octupole--octupole interactions. Preliminary results show that they become important for $f_{wt}$ larger than a few percent ($L$ lower than, say, 10). The interactions introduce local correlations which modify the local and global order. This study is under progress and its results to be published elsewhere.

\appendix

\section{Quantum-Chemical validation of the model}

The choice of point-charge model for octupolar molecules implies that interactions with an external electric field  are of purely electrostatic nature. However, quantum effects can, potentially, introduce corrections. In this Section we use some simple methods of QChem to address this issue.

\subsection{Octupole moment} \label{s:OctupoleMoment}

Contributions to the octupole moment come from both electrons and  nuclear charges. The electronic part was calculated using DFT in the \emph{Dalton} program.\cite{daltonpaper,ref:dalton} The electronic charge density, obtained from the solution of Kohn--Sham  equations,\cite{PielaQChemBook} was numerically integrated to calculate the Cartesian components of the electronic contribution to the octupole moment, see Eq.\ \eqref{eq:octupoleCartDef}. The calculations used the DFT hybrid functional B3LYP and cc--pVDZ basis set. All calculations in this Appendix were done for a planar geometry of the TATB molecule optimized with the use of the B3LYP/6--31G(d) method. The magnitude of the total octupole moment (including electronic and nuclear contribution) was found to be
\begin{equation}
\norm{\mathcal{O}}\approx 
 28.8\; \eAt. \label{eq:OctupoleMomentDalton}
\end{equation}

In the point-charge octupole model, Eq.\ \eqref{eq:octupoleNorm}, the octupole moment is specified by two parameters, $q$ and $d$. In this paper we use $q=0.66\;e,\, d=2.44$~\AA. Other combinations (e.g.\ $q=0.5\;e,\, d=2.67$~\AA\ ) have negligible impact on quantitative results. The parameters of a typical octupole used in Ref.~\onlinecite{MitusPZ1} yield the value $||\mathcal{O}||=6\;\eAt$. The calculated octupole moment of TATB is thus nearly five times larger, which increases the energy barrier accordingly (see Eq.\ \eqref{eq:barrier6QOct}).

\subsection{Energy barrier: quantum corrections} \label{s:TestEnergyBarrier}

\begin{table}
	\centering
	\begin{tabular}{c|cccc}
		\hline\hline
		Q [$e$] & 1 & 8 & 1/2 & 1 \\
		R [nm] & 4 & 4 & 2 & 2 \\ \hline
		$\DE_{\rm QChem}/\kB$ [K] & 28.4 & 227 & 231 & 462 \\
		$\DE/\kB$ [K]  & 28.2 & 226 & 227 & 454 \\
	\hline\hline
    \end{tabular}
	\caption{Energy barriers computed by QChem methods ($\DE_{\rm QChem}/\kB$) and $\DE/\kB$ predicted by the classic 6AO model.
	}
	\label{tab:QChem6QBarrier}
\end{table}

The energy barrier for the octupolar TATB molecule at the center of the poling cell  (see Fig.~\ref{fig:poling-cell-charges})  was roughly estimated using QChem methods and compared to the barrier predicted in the  point charges poling scheme. The QChem computations of energy of TATB molecule were done at the B3LYP/cc--pVDZ level, without optimalization of the molecular structure.  The classic formula for the energy barrier of a model 6AO  is given by Eq.\ \eqref{eq:barrier6QOct}. We have found that both barriers have similar values even for much stronger poling fields  than those used in this study, represented by four sets of values of $Q$ and $R$  in Table~\ref{tab:QChem6QBarrier}.

\bibliographystyle{aip}
\bibliography{2d-nowy}

\begin{thebibliography}{10}

\bibitem{BoydNLO}
R.~W. Boyd,
\newblock {\em Nonlinear Optics},
\newblock Academic Press, 2nd edition, 2003.

\bibitem{KielichNLO}
S.~Kielich,
\newblock {\em Molekularna Optyka Nieliniowa {\rm (in polish)}},
\newblock PWN, Warszawa, 1977.

\bibitem{chemla1987nonlinear}
D.~S. Chemla and J.~Zyss, editors,
\newblock {\em Nonlinear Optical Properties of Organic Molecules and Crystals},
\newblock Number t. 1 in Materials Science and Technology Series, Academic
  Press, 1987.

\bibitem{MaroulisAtomicmolecularnonlinear2011}
G.~Maroulis, T.~Bancewicz, and B.~Champagne,
\newblock {\em Atomic and molecular nonlinear optics: {Theory}, experiment and
  computation: {A} homage to the pioneering work of {S}tanis{\l }aw {K}ielich
  (1925-1993)},
\newblock IOS Press, 2011.

\bibitem{messier2012organic}
J.~Messier, F.~Kajzar, and P.~Prasad,
\newblock {\em Organic molecules for nonlinear optics and photonics},
\newblock Springer Netherlands, 2012.

\bibitem{LafargueLocalizedlasingmodes2014}
C.~Lafargue et~al.,
\newblock Phys. Rev. E {\bf 90}, 052922 (2014).

\bibitem{HajjElectroopticalPockelsscattering2011}
B.~Hajj et~al.,
\newblock Opt. Express {\bf 19}, 9000 (2011).

\bibitem{CastagnaNanoscalePolingPolymer2013}
R.~Castagna, A.~Milner, J.~Zyss, and Y.~Prior,
\newblock Advanced Materials {\bf 25}, 2234 (2013).

\bibitem{BrasseletNanoCrystalsQuadraticNonlinear2010}
S.~Brasselet and J.~Zyss,
\newblock Nano-{Crystals} for {Quadratic} {Nonlinear} {Imaging}:
  {Characterization} and {Applications},
\newblock in {\em Nanocrystals}, edited by Y.~Masuda, page~24, INTECH Open
  Access Publisher, 2010.

\bibitem{PeyronelQuantumnonlinearoptics2012a}
T.~Peyronel et~al.,
\newblock Nature {\bf 488}, 57 (2012).

\bibitem{ZyssRelationsmicroscopicmacroscopic1982}
J.~Zyss and J.~L. Oudar,
\newblock Physical Review A {\bf 26}, 2028 (1982).

\bibitem{ZyssChiralityhydrogenbonding1984}
J.~Zyss, J.~F. Nicoud, and M.~Coquillay,
\newblock The Journal of Chemical Physics {\bf 81}, 4160 (1984).

\bibitem{ZyssFirst}
J.~Zyss,
\newblock Nonl.~Opt. {\bf 1}, 3 (1991).

\bibitem{ZyssJCP93engineeringImplications}
J.~Zyss,
\newblock J.~Chem.~Phys. {\bf 98}, 6583 (1993).

\bibitem{BZ98}
S.~Brasselet and J.~Zyss,
\newblock J. Opt. Soc. Am. B {\bf 15}, 257 (1998).

\bibitem{MitusPZ1}
A.~C. Mitu\'{s}, G.~Pawlik, and J.~Zyss,
\newblock J.~Chem.~Phys. {\bf 135}, 024110 (2011).

\bibitem{ACMNloQo}
A.~C. Mitu\'{s}, M.~Jarema, G.~Pawlik, and J.~Zyss,
\newblock Nonl.~Opt.,~Quant.~Opt. {\bf 43}, 133 (2012).

\bibitem{ActaPhysPolB}
M.~Jarema, A.~C. Mitu\'{s}, and J.~Zyss,
\newblock Acta~Phys.~Pol.~B {\bf 43}, 1017 (2012).

\bibitem{TATB-SHG-1990}
I.~Ledoux, J.~Zyss, J.~S. Siegel, J.~Brienne, and J.-M. Lehn,
\newblock Chem.~Phys.~Lett. {\bf 172}, 440  (1990).

\bibitem{ExtededDipolesACMGPIRFK08}
A.~C. Mitu\'{s}, G.~Pawlik, I.~Rau, and F.~Kajzar,
\newblock Nonl.~Opt.,~Quant.~Opt. {\bf 38}, 141 (2008).

\bibitem{MJPhD}
M.~Jarema,
\newblock \emph{PhD Thesis}, Wroc\l{}aw University of Science and Technology
  (2015).

\bibitem{Jerphagnon}
J.~Jerphagnon, D.~Chemla, and R.~Bonneville,
\newblock Adv. in Phys. {\bf 27}, 609 (1978).

\bibitem{PatashinskiNanofluidicManifestationsStructure2019}
A.~Z. Patashinski, M.~A. Ratner, R.~Orlik, and A.~C. Mitus,
\newblock J. Phys. Chem. C {\bf 123}, 16787 (2019).

\bibitem{BesselWatson}
G.~N. Watson,
\newblock {\em A Treatise on the Theory of Bessel Functions},
\newblock Cambridge University Press, Cambridge, 1996.

\bibitem{LevineSecondthirdorder1975}
B.~F. Levine and C.~G. Bethea,
\newblock J.~Chem.~Phys. {\bf 63}, 2666 (1975).

\bibitem{OudarOpticalnonlinearitiesconjugated1977}
J.~L. Oudar,
\newblock J.~Chem.~Phys. {\bf 67}, 446 (1977).

\bibitem{DipoleMoments}
A.~L. McClellan,
\newblock {\em Table of Experimental Dipole Moments},
\newblock Freeman, San Francisco, 1963.

\bibitem{Sinhagroundexcitedstate1991}
H.~K. Sinha and K.~Yates,
\newblock Canadian Journal of Chemistry {\bf 69}, 550 (1991).

\bibitem{Note1}
For simplicity, we neglected the local-field corrections, which have similar
  effect on dipoling and octupoling, and are not important for the
  order-of-magnitude estimations.

\bibitem{ZyssAMARIS}
J.~Zyss,
\newblock Nonl.~Opt.,~Quant.~Opt. {\bf 43}, 97 (2012).

\bibitem{GunterNLOMaterialsBook}
P.~G\"{u}nter, editor,
\newblock {\em Nonlinear optical effects and materials},
\newblock Springer, Berlin, 2000.

\bibitem{PolymerDielectricStrength}
See e.g.
  http://polymerdatabase.com/polymer\%20physics/\linebreak[0]Dielectric\%20Strength.html,
  access date: 30.06.2020.

\bibitem{PolymerBook}
L.~A. Dissado and J.~C. Fothergill,
\newblock {\em Electrical Degradation and Breakdown in Polymers},
\newblock Peter Peregrinus Ltd., London, 1992.

\bibitem{GroteDNACTMADielectricStrength}
R.~A. Norwood et~al.,
\newblock Proc. SPIE {\bf 7403}, 74030A (2009).

\bibitem{Diamond1983SovPhys}
E.~Konorova et~al.,
\newblock Soviet physics. Semiconductors {\bf 17}, 146 (1983).

\bibitem{DiamondElectronicMat}
C.~J. Wort and R.~S. Balmer,
\newblock Materials Today {\bf 11}, 22  (2008).

\bibitem{DiamondLuminescence5e8V-m}
T.~Watanabe et~al.,
\newblock Japanese Journal of Applied Physics, Part 2: Letters {\bf 40}, L715
  (2001).

\bibitem{Diamond-3e9}
M.~Irie, S.~Endo, C.~Wang, and T.~Ito,
\newblock Diamond and Related Materials {\bf 12}, 1563 (2003).

\bibitem{PMMA-tensileStrength}
See, e.g. https://www.ipolymer.com/pdf/\linebreak[0]Acrylic.pdf,
  https://\linebreak[0]designerdata.nl/materials/\linebreak[0]plastics/thermo-plastics/\linebreak[0]poly(methyl-methacrylate),
  access date: 30.06.2020.

\bibitem{DielectricStrengthSiO2}
See e.g. http://accuratus.com/fused.html, access date: 30.06.2020.

\bibitem{DNDABreport}
I.~Ledoux, I.~Cazenobe, S.~Brasselet, E.~Toussaere, and J.~Zyss,
\newblock Quantum Electronics and Laser Science Conference, {\rm San Francisco}
   (2000).

\bibitem{IreneCazenobePhD}
I.~Cazenobe,
\newblock \emph{PhD Thesis}, University of Paris XI  (2015),
\newblock http://www.theses.fr/1999PA112153, access date: 08.2020.

\bibitem{CRCDipole}
D.~R. Lide, editor,
\newblock {\em CRC Handbook of Chemistry and Physics},
\newblock CRC Press, 84th edition, 2003.

\bibitem{YaronOproptoB}
J.~D. Weibel, D.~Yaron, and J.~Zyss,
\newblock J.~Chem.~Phys. {\bf 119}, 11847 (2003).

\bibitem{JACS1992}
J.~L. Bredas, F.~Meyers, B.~M. Pierce, and J.~Zyss,
\newblock J. Am. Chem. Soc. {\bf 114}, 4928 (1992).

\bibitem{JeongOctupolarMoleculesNonlinear2015}
M.-Y. Jeong and B.~R. Cho,
\newblock The Chemical Record {\bf 15}, 132 (2015).

\bibitem{daltonpaper}
K.~Aidas et~al.,
\newblock WIREs Comput.~Mol.~Sci. {\bf 4}, 269 (2014).

\bibitem{ref:dalton}
\textit{Dalton}, a molecular electronic structure program, Release v2011
  (2011), see http://daltonprogram.org.

\bibitem{PielaQChemBook}
L.~Piela,
\newblock {\em Ideas of quantum chemistry},
\newblock Elsevier, Amsterdam, 2nd edition, 2014.

\end{thebibliography}

\end{document}